\documentclass[aps,twocolumn,showpacs,preprintnumbers,noshowpacs,superscriptaddress,6pt]{revtex4-1}
\usepackage[breaklinks]{hyperref}

\usepackage[labelfont=bf]{caption}
\usepackage{euscript}
\usepackage{epsf}
\usepackage{epsfig}
\usepackage{subcaption}

\usepackage[utf8]{inputenc}
\usepackage{graphicx,epstopdf}
\usepackage{overpic}
\usepackage{multirow}
\usepackage{adjustbox}
\usepackage{natbib}

\usepackage{parskip}
\usepackage[T1]{fontenc}
\usepackage{dcolumn}

\usepackage{bm}
\usepackage{amsfonts}
\usepackage{amsmath}
\usepackage{amssymb}
\usepackage{siunitx}
\usepackage{booktabs}
\usepackage{mathtools}
\usepackage{flafter}
\usepackage{threeparttable}
\usepackage{subfiles}
\usepackage{import}
\usepackage{float}
\usepackage{makecell}
\usepackage{diagbox}
\usepackage{indentfirst}
\usepackage{mdframed}
\usepackage[strict]{chngpage}

\newmdenv{allfour}

\newmdenv[leftline=false,rightline=false]{topbot}

\newmdenv[topline=false,rightline=false]{leftbot}

\usepackage{ulem}

\usepackage{color, colortbl}

\definecolor{Gray}{gray}{0.9}
\definecolor{LightCyan}{rgb}{0.88,1,1}

\definecolor{BLUE}{rgb}{0.0,0.0,1.0}

\graphicspath{{images/}}

\everymath{\displaystyle}
\renewcommand{\vec}[1]{{\mbox{\boldmath$#1$}}}

\begin{document}

\title{
The ground-state potential and dipole moment of carbon monoxide: contributions from electronic correlation, relativistic effects, QED, adiabatic, and non-adiabatic corrections
}

\author{D.~P.~Usov}
\affiliation{Department of Physics, St. Petersburg State University, 7/9 Universitetskaya nab., 199034 St. Petersburg, Russia}

\author{Y.~S.~Kozhedub}
\email[]{y.kozhedub@spbu.ru}
\affiliation{Department of Physics, St. Petersburg State University, 7/9 Universitetskaya nab., 199034 St. Petersburg, Russia}

\author{V.~V.~Meshkov}
\affiliation{Department of Chemistry, Lomonosov Moscow State University, 1/3 Leninskie gory, 119991 Moscow, Russia}

\author{A.~V.~Stolyarov}
\affiliation{Department of Chemistry, Lomonosov Moscow State University, 1/3 Leninskie gory, 119991 Moscow, Russia}

\author{N.~K.~Dulaev}
\affiliation{Department of Physics, St. Petersburg State University, 7/9 Universitetskaya nab., 199034 St. Petersburg, Russia}
\affiliation{National Research Centre “Kurchatov Institute” B.P. Konstantinov Petersburg Nuclear Physics Institute, Gatchina,
Leningrad district 188300, Russia}

\author{N.~S.~Mosyagin}
\affiliation{National Research Centre “Kurchatov Institute” B.P. Konstantinov Petersburg Nuclear Physics Institute, Gatchina,
Leningrad district 188300, Russia}

\author{A.~M.~Ryzhkov}
\affiliation{Department of Physics, St. Petersburg State University, 7/9 Universitetskaya nab., 199034 St. Petersburg, Russia}
\affiliation{National Research Centre “Kurchatov Institute” B.P. Konstantinov Petersburg Nuclear Physics Institute, Gatchina,
Leningrad district 188300, Russia}

\author{I.~M.~Savelyev}
\affiliation{Department of Physics, St. Petersburg State University, 7/9 Universitetskaya nab., 199034 St. Petersburg, Russia}

\author{V.~M.~Shabaev}
\affiliation{Department of Physics, St. Petersburg State University, 7/9 Universitetskaya nab., 199034 St. Petersburg, Russia}
\affiliation{National Research Centre “Kurchatov Institute” B.P. Konstantinov Petersburg Nuclear Physics Institute, Gatchina,
Leningrad district 188300, Russia}

\author{I.~I.~Tupitsyn}
\affiliation{Department of Physics, St. Petersburg State University, 7/9 Universitetskaya nab., 199034 St. Petersburg, Russia}

\date{\today}

\begin{abstract}
The ground X$^1\Sigma^+$ state potential energy curve (PEC) and dipole moment curve (DMC) of CO molecule have been revisited within the framework of the relativistic coupled-cluster approach, which incorporates non-perturbative single, double, and triple cluster amplitudes (CCSDT) in conjunction with a finite-field methodology. The generalized relativistic pseudo-potential model was used for the effective introducing the relativity in all-electron correlation treatment and accounting the quantum-electrodynamics (QED) corrections within the model-QED-operator approach. The diagonal Born-Oppenheimer correction to PEC has been evaluated using the CCSD approach. The sensitivity of resulting PEC and DMC to variations in basis set parameters and regular intramolecular perturbations were considered as well. The present \textit{ab initio} results are in a reasonable agreement with their most accurate semi-empirical counterparts. 
\end{abstract}

\maketitle
\section{Introduction}\label{sec:intro} 
Carbon monoxide (CO) stands out as one of the most resilient diatomic molecules in the Universe, owing to its robust 'triple' chemical bond. 
Indeed, the CO($J=2\to 1$) rotational emission is observed in Early Universe at high red-shift $z\sim 5-7$~\cite{Riechers2019}. 
In fact, next to molecular hydrogen, CO ranks among the most prevalent diatomic species~\cite{Combes:1991}. Spectral signatures of CO have been detected in diverse celestial settings, including solar~\cite{Lyons:2018} and stellar~\cite{Abia:2012} atmospheres, and have even been observed on celestial bodies such as Mars~\cite{Olsen:2021} and Venus~\cite{Vandaele:2016}. Contemporary studies, employing methods like cross-correlation~\cite{Giacobbe:2021}, are identifying CO spectral lines in exoplanet's atmospheres. In addition to its cosmic presence, CO is a notable terrestrial pollutant, particularly in the troposphere, stemming from both natural sources like biomass combustion and anthropogenic origins, such as automobile emissions~\cite{Beale:2016}. Consequently, an exhaustive understanding of CO's spectral characteristics is imperative for precise modeling and monitoring of a broad spectrum of astrophysical, environmental, and atmospheric phenomena (see, for instance, the HITRAN molecular spectroscopic database~\cite{GORDON2022}).
\par
Moreover, the carbon monoxide molecule serves as a crucial benchmark system for high-accuracy studies of absorption line intensities~\cite{PRL2022,Balashov:JCP:2023}. Experimentally, CO offers distinct advantages, given its ease of concentration determination and stability. The diatomic nature of the molecule and its relatively large reduced mass ensures well-isolated absorption rotational lines for most abundant isotopologues. 
{\it Ab initio} electronic structure calculations of CO benefit from relatively small number of electrons to be correlated explicitly~\cite{Zaitsevskii2021} and the absence of light (H) atoms, diminishing mass-dependant non-adiabatic effects that dominate in more light diatomic systems, such as molecular hydrogen and hydrids.
\par
Given its key role in numerous domains, there has been a wealth of experimental and theoretical investigations dedicated to CO spectroscopy over the past century, e.g., Refs.~ \cite{Chackerian:1982, Ermilov:1990, Coxon:JCP:2004} to mention a few.
Recently, a series of theoretical studies has been conducted to investigate the ground-state potential and dipole moment of carbon monoxide~\cite{PRL2022, Peng_Fei:CTP:2013,Chen:CPB:2015,Choluj:CPL:2016,Konovalova:OS:2018,Ushakov:PCCP:2020,Lykhin:JCTC:2021,Balashov:JCP:2023,Meshkov:JQSRT:2022,Meshkov:RJPC:2023,Dulaev:OS:2023}. However, the extensive dataset available may not be considered exhaustive due to its specificity or lack of systematic coverage.
In order to provide a precise theoretical description of the potential energy levels and various properties of CO at a modern level of accuracy, careful consideration of correlation and relativistic effects is required. Despite the molecule being relatively light, the contributions from relativistic and potentially quantum electrodynamics (QED) corrections prove to be significant. Furthermore, this molecule was proposed for studying the variation of the fine-structure constant~\cite{Konovalova:OS:2018, Zaitsevskii2021}. It is also worth noting the high sensitivity of the intensity distribution in the rovibrational spectrum to a permanent dipole moment curve of the molecule~\cite{gordon2016}.

The primary objective of this study is to present a comprehensive and contemporary {\it ab initio} evaluation of ground state potential energy curve (PEC) and dipole moment curve (DMC) for carbon monoxide with the utmost precision.
The investigation employs the exhausted approach rooted in contemporary quantum theory to accurately address correlation, relativistic and QED effects as well as adiabatic correction and a rotational part of non-adiabatic correction within the ground-state of molecule. Specifically, we adopt the relativistic coupled-cluster method encompassing nonperturbative single (S), double (D), and, notably, entirely triple (T) cluster amplitudes (CCSDT). These computations are conducted employing the all-electron Dirac-Coulomb (DC) Hamiltonian as well as its two-component analogue (X2C)~\cite{x2cmmf}.
The assessment of the Breit and QED corrections is conducted utilizing the framework of generalized relativistic pseudo-potentials (GRPP)~\cite{GRPP,Oleynichenko:Sym:2023}.
 Incorporation of the QED contribution is based on the paradigm of the model-QED-operator approach, as exemplified in references~\cite{QEDMOD1,Shabaev:CPC:2018}.
 
A challenging aspect inherent to theoretical explorations within this domain is the absence of a straightforward methodology for the robust evaluation of uncertainties.
Nonetheless, as it has been recently demonstrated by fine {\it ab initio} calculation of rovibrational intensities in a very weak (0-7) overtone~\cite{Balashov:JCP:2023} that the estimation of accuracy remains essential to guarantee reliable prediction of carbon monoxide  properties, as well as interpretation of experimental data. In this context, uncertainties derived from pure theoretical analyses are typically given preeminence. A methodology that exhibits transparency, both in terms of the approximations employed and the considered effects, while also facilitating systematic enhancements, constitutes a essential for a reliable theoretical uncertainty assessment.
\par
The manuscript is organized as follows. Section~\ref{sec:methods} provides a concise exposition of our methodological approaches, accompanied by a short overview of their implementation details. In Section~\ref{sec:results}, an exhaustive description of the calculation methodology is presented. This section further engages in a comprehensive discussion of the obtained outcomes, together with a comparative analysis with existing literature values.
%
\section{Theoretical approaches and methods}\label{sec:methods}
%
\subsection{Accounting for relativistic and quantum-electrodynamics corrections}\label{RelQDT}

The calculations are performed within the framework of the relativistic four-component Dirac–Coulomb (DC) Hamiltonian, employing the computational software package DIRAC~\cite{DIRAC19, Saue:2020} and its relevant extensions. A diatomic molecule is considered within the Born-Oppenheimer approximation, where the nuclei are the sources of electrostatic potential, fixed at the positions $\vec{R}_{1, 2}$ with the internuclear distance $R=|\vec{R}_{2}-\vec{R}_{1}|$.
The DC Hamiltonian of the electronic system (in atomic units) is given by
\begin{equation}
\label{eq:HDC}
    H_{\text{DC}} = \Lambda^+ \left[\sum_i c \vec{\alpha}_i \cdot \vec{p}_i + \beta_i c^2 + V(\vec{r}_i) + \sum_{i < j} \frac{1}{ |\vec{r}_i - \vec{r}_j|} \right] \Lambda^+,
\end{equation}
where $ \beta $ and $ \vec{\alpha} $ are the standard $ 4 \times 4 $ Dirac matrices, $\vec{r}_i$ and $\vec{p}_i$ are the position vector and momentum of the $i$-th electron, $ V(\vec{r}_i) = V^{(1)}_{\text{nucl}}(|\vec{R}_1 - \vec{r}_i|) + V^{(2)}_{\text{nucl}}(|\vec{R}_2 - \vec{r}_i|) $ is the total nuclear binding potential (the finite nuclear size effect is taken into account within the Gaussian model of charge distribution), the summation goes over all electrons of the system, and $ \Lambda^+ $ is the projector on the positive-energy Dirac-Fock one-electron states. 
\par
There also exists the so-called eXact-2-Component (X2C) Hamiltonian, which serves as an approximation to the DC Hamiltonian $H_{\text{DC}}$ \eqref{eq:HDC}, yet precisely replicates the positive-energy spectrum of the four-component one-electron Hamiltonian~\cite{Kutzelnigg:2005, Ilias:2007}. Utilizing a 2-component framework, this Hamiltonian notably offers enhanced numerical efficiency.
Furthermore, within the computational framework of the DIRAC package, the X2C Hamiltonian stands as the exclusive option that facilitates the incorporation of the Gaunt interaction as spin-same and spin-other orbit mean-field operators within the context of the Atomic Mean-Field Integral (AMFI) approximation. A comprehensive exploration of this aspect is available in Ref.~\cite{Ilias:2001} and the associated references.
The Gaunt term, denoted as~$V^{\mathrm{G}}$ 
\begin{equation}
V_{ij}^\mathrm{G} = - \frac{(\bm{\alpha}_i \cdot \bm{\alpha}_j)}{r_{ij}},
\end{equation}
constitutes the foremost relativistic correction to the Coulomb interaction. 
In this study, we employ the molecular mean-ﬁeld X2C modification~\cite{x2cmmf}.
\par

Another optimization used in the present work is based on calculations with the generalized relativistic pseudopotentials (GRPP)~\cite{GRPP,Oleynichenko:Sym:2023}.
This method has proven to be highly effective as it allows for the description of interactions with internally "core"\, electrons that are excluded from the calculation, accurately accounting for relativistic corrections, Breit interaction, and other corrections without performing a full four-component calculation. In this study, a nonlocal formulation of the GRPP with a zero core was employed. In addition to relativistic corrections and the Breit interaction, the QED correction was included in the GRPP potential~\cite{Meshkov:RJPC:2023}.
Describing the QED effects is based on the model-QED-operator approach~\cite{Shabaev:CPC:2015,Shabaev:CPC:2018}.
This single-electron operator is independently constructed for each atom of the molecule as a sum of the vacuum polarization operator and the self-energy operator. The vacuum polarization operator is described by a sum of local Uehling and Wichmann-Kroll potentials, while the self-energy operator is approximated by a sum of short-range quasi-local and nonlocal potentials. The molecular QED operator is represented as a superposition of atomic operators. This approximation is justified as the QED operator has a small range of action and is concentrated in the internal "core"\, region of the atoms, while we are interested in the range of internuclear distances \mbox{$R> 0.6$~\AA}, which is far enough from the United Atom and, hence, responsible for the formation of the so-called "chemical"\, bond in the molecule.
The calculations were performed using the DIRAC program and the LIBGRPP library~\cite{Oleynichenko:Sym:2023}, which is necessary for accounting for the nonlocal part of the potential.

\subsection{Electronic correlation treatment}\label{CCSDT}
\par
Utilizing the frameworks of DC, X2C, or GRPP Hamiltonians, the electronic structure is addressed through the application of the single-reference relativistic coupled cluster (CC) methodology.
For the current investigation, we implement the CC method as realized in the EXP-T program~\cite{Oleynichenko:website,Oleynichenko:2020}.
In our paper, we adopt the designation CCSD for calculations incorporating the single (S) and double (D) cluster amplitudes, while CCSDT signifies calculations involving both SD and nonperturbative triple (T) amplitudes. Instances where the triple amplitudes are perturbatively evaluated are identified as CCSD(T).
The Dirac-Hartree-Fock (DHF) calculations, alternatively known as relativistic Hartree-Fock computations, alongside the subsequent integral transformation procedures, are carried out utilizing the DIRAC package.

The CC calculations are performed across diverse configurations, involving varying numbers of correlated electrons and virtual orbitals.
For the conclusive version of the computations, the standard basis sets from the cc-pV$N$Z family ($N = 3, 4, 5, 6$)~\cite{Dunning:JCP:1989,Kendall:JCP:1992} are employed.

\subsection{Extrapolation to the complete basis set}\label{CBS}
A comprehensive exploration of these calculations entails augmenting the aforementioned basis sets through the sensible addition of further diffuse (low exponent) basis functions, incorporated in an even-tempered manner. To address the inherent limitations of basis-set completeness, an extrapolation approach to attain the complete basis set (CBS) limit is employed.
For the outcomes stemming from the Dirac-Hartree-Fock methodology, the extrapolation scheme detailed below is adopted:
\begin{equation}
E_{\rm DHF}(N) = E_{\rm DHF}^{\rm CBS} + Ae^{-\beta N}.
\end{equation}
In the context of correlation corrections, the following formula is applied: 
\begin{equation}
E_{\rm corr}(N) = E_{\rm corr}^{\rm CBS} + \frac{A}{N^3}.
\end{equation}
The rationale underpinning of the utilization of this extrapolation technique within molecular computations involving correlation-consistent basis sets is expounded upon in the work referenced as~\cite{EXTRAPOLATION}.
In the pursuit of methodological robustness, Dyall’s relativistic basis sets~\cite{Dyall:2009, Gomes:2010,Dyall:2016,DYALL}, varying in quality and employed in an uncontracted manner, are also incorporated within the realm of test calculations.

\subsection{The finite-field calculation of DMC}\label{FFDMF}
The dipole moment $\vec{d}$ is computed employing the finite-field (FF) methodology. This involves the expansion of the energy of the molecule subjected to a weak and uniform electrostatic field $\vec{F} = (F_x, F_y, F_z)$ through the Taylor series:
\begin{equation}\label{eq:ff}
    E(\vec{F}) = E_0 - \sum_i d_i F_i + \dots
\end{equation}
Subsequently, the components of $\vec{d}$ are defined from this series~{\eqref{eq:ff}} as
\begin{equation}\label{eq:ff-derivs}
    d_i = -\left. \frac{\partial E(\vec{F})}{\partial F_i} \right\vert_{\vec{F} = 0}, 
\end{equation}
where $i = x, y, z$.
The numerical determination of derivatives, as given in equation~\eqref{eq:ff-derivs}, is achieved through the energies $E(\vec{F})$ obtained across multiple instances of $\vec{F}$.
In principle, the finite-field approach stands as an exact method. Nevertheless, a judicious selection of the field strength is imperative to ensure precise numerical differentiation. Given the axial symmetry characterizing the studied diatomic molecule, solely the components of $\vec{d}$, aligned with the molecular axis, exhibit nontrivial values.

\subsection{Evaluation of adiabatic correction to PEC}\label{DBOC}

The adiabatic correction to the Born-Oppenheimer (BO) approximation implemented in the quantum chemical treatment above can be easily computed using perturbation theory (PT). The diagonal correction to the ground-state PEC is obtained in 1-st order and takes the form
\begin{equation}\label{eq:DBOC}
    U_{\rm ad} = \sum_{I={\rm C},{\rm O}} -\frac{\hbar^2}{2M_I}\langle \Psi^{\rm el}_X|\Delta_I |\Psi^{\rm el}_X\rangle, 
\end{equation}
where $M_I$ are the nuclear mass of C and O atoms, $\Psi^{\rm el}_X$ is the electronic wave function of the ground $X$-state depending on the internuclear distance $R$ as a parameter, $\Delta_I$ is the Laplace operator. Differentiations in Eq.~(\ref{eq:DBOC}) are over nuclear Cartesian coordinates while the integration is over electronic coordinates.

The mass-dependent function $U_{\rm ad}(R)$, commonly referred to as the "diagonal Born-Oppenheimer correction" (DBOC), introduces variations in the effective interatomic PECs for different isotopologues of a molecule~\cite{PEF_EMPIRICAL, Coxon:JCP:2004}. 
The adiabatic correction to the $X$-state PEC, specifically for the primary isotopologue $^{12}$C$^{16}$O, was assessed using analytic derivative techniques. This computational approach was implemented in the CFOUR program package~\cite{cfour}. The calculations were performed at the non-relativistic Hartree-Fock (HF) and CCSD levels, employing the aug-cc-pCV$N$Z ($N$=3,4) basis sets. Notably, two lowest molecular orbitals were kept frozen during the CCSD calculations.

\subsection{Propagation of the intramolecular perturbations into the PEC and DMC}\label{BOB}

The energy-isolated ground state of CO molecule still undergoes the regular intramolecular interactions with the higher-lying singlet and triplet states~\cite{field}. The impact of these very weak perturbations on energy of the $X$-state can be estimated in the framework of the $2$-nd order PT:
\begin{eqnarray}\label{dcorr0}
\delta E_X \approx \sum_{j}\sum_{v_j}\frac{|V^{\rm pert}_{v_Xv_j}|^2}{E_{v_X}-E_{v_j}}
\end{eqnarray}
as the sum over an infinite number of the bound and embedding in continuum vibronic states. The tedious summation and integration over vibrational $v_j$-states can be avoided due to the approximated vibrational sum rule~\cite{pupyshev1994, pazyuk1994}. Then, the regular perturbation effect of the remote states manifold on the BO PEC and DMC can be represented as 
\begin{eqnarray}\label{dcorr1}
\delta U_X \approx \sum_{j}\frac{|V^{\rm pert}_{Xj}|^2}{\Delta U_{Xj}};
\quad 
\delta d_X \approx 2\sum_{j}\frac{V^{\rm pert}_{Xj}d_{Xj}}{\Delta U_{Xj}}
\end{eqnarray}
where $\Delta U_{Xj}(R)=U^{\rm BO}_X - U^{\rm BO}_j$ is the difference of the BO potentials, $d_{Xj}(R)$ is the spin-allowed electronic transition dipole moment between the ground and excited singlet states, whereas $V^{\rm pert}_{Xj}(R)$ is the relevant non-adiabatic electronic matrix element assumed to be a multiplicative function of $R$. The summation in Eq.~(\ref{dcorr1}) should be performed over all excited states. However, under {\it unique perturber} approximation~\cite{field} one can select a single electronic state giving the dominant contribution to the sum.

In the case of {\it homogeneous} perturbations, which obey $\Delta\Lambda=0$ selection rule, the non-adiabatic coupling matrix elements $V^{\rm pert}_{Xj}$ are mainly determined by the mass-invariant electrostatic interaction between the states of the same $j\in~ 
 ^1\Sigma^+$ symmetry. These matrix elements are unambiguously related to the non-Hermitian radial coupling matrix elements $B_{Xj}(R)=\langle \Psi^{\rm el}_X|\partial/\partial R |\Psi^{\rm el}_j\rangle$ which can be evaluated using {\it ab initio} methods~\cite{MOLPRO}. However, their transformation to the multiplicative Hermitian function $V^{\rm pert}_{Xj}(R)$ is not straightforward.

In the case of {\it heterogeneous} perturbations fulfilling the selection rule $\Delta\Lambda=\pm 1$, the function is
\begin{eqnarray}\label{het}
V^{\rm pert}_{Xj}=-BL_{Xj}\sqrt{2J(J+1)}; 
\quad B\equiv\left(\frac{\hbar^2}{2\mu R^2}\right )
\end{eqnarray}
where $\mu=\frac{M_{\rm C}M_{\rm O}}{M_{\rm C}+M_{\rm O}}$ is the reduced molecular mass, $J$ is the rotational quantum number, and $L_{Xj}(R)=\langle \Psi^{\rm el}_X|\hat{L}_x\pm i\hat{L}_y|\Psi^{\rm el}_j\rangle/\sqrt{2}$ is the $L$-uncoupling matrix element of electronic angular momentum operator responsible for the so-called "Coriolis" electronic-rotational interaction.

Then, inserting of the operator (\ref{het}) into the PT relations (\ref{dcorr1}) leads to the diagonal $J$-dependent correction to the BO PEC:
\begin{eqnarray}\label{hetU}
\delta U_X \approx B\left[1+q\right]J(J+1);\quad 
q = 2B\left [\sum_{j}\frac{|L_{Xj}|^2}{\Delta U_{Xj}}\right]
\end{eqnarray}
and BO DMC
\begin{eqnarray}\label{hetd}
\delta d_X \approx -2B\left[\sum_{j}\frac{L_{Xj}d_{Xj}}{\Delta U_{Xj}}\right]\sqrt{2J(J+1)}
\end{eqnarray}
where $d_{Xj}(R)=\langle \Psi^{\rm el}_X|\hat{d}_x\pm i\hat{d}_y|\Psi^{\rm el}_j\rangle/\sqrt{2}$ is the electronic transition dipole moment between the ground and excited $j\in~^1\Pi$ state. 

The summations over the upper states in Eqs.~(\ref{hetU}) and~(\ref{hetd}) were restricted in the present work by the lowest $j\in (1-3)^1\Pi$ terms. The required electronic $L_{Xj}$ and $d_{Xj}$ matrix elements, along with the relevant BO PECs, were obtained within the framework of the internally contracted multi-reference configuration interaction (ic-MR-CI) calculations implemented in the MOLPRO package~\cite{MOLPRO}. Both state-averaged CASSCF and ic-MR-CI calculations have been accomplished with the aug-cc-pCVQZ-DK basis set in the $6/2/2/0$ active space while two lowest orbitals were frozen.

\section{Results and discussions }\label{sec:results}

Evaluations of both BO PEC and DMC for the ground $X^1\Sigma^+$ state of carbon monoxide have been performed in the range of $R$ from $0.6$ to $1.6$~\AA, which is restricted to the convergence region of the single-reference CC methods used. For DMC calculations within the finite-field scheme the optimal value of the electric field strength $F=\pm0.0001$~a.u. is found suitable enough for accurate numerical differentiation.

To undertake a comprehensive exploration of correlation and relativistic influences, our study encompasses various facets of the calculations and associated approximations:
\begin{enumerate}
\item Hamiltonian Comparisons: A comparative analysis is performed between the DC and X2C as well as GRPP Hamiltonians to discern their respective impacts;
\item Relativistic and QED Considerations: Both relativistic and QED contributions are meticulously examined.
\item Correlation Space Size: We scrutinize the dimensionality of the correlation space, encompassing considerations, such as the total count of explicitly correlated electrons and the energy threshold utilized for virtual orbitals within the framework of the CC scheme.
\item Basis Set Completeness: The effect of finite one-electron basis set completeness, inclusive of diffuse orbital basis functions, is assessed.
\item Higher-Order Excitations: A dedicated analysis of the influence arising from triple excitations is conducted, aiming to gauge their relative contributions.
\end{enumerate}

In such a way the results of calculations obtained within the DC Hamiltonian and X2C approximation are almost identical. That is why using the last one is fully justified for relativistic study of the molecule under consideration. 

\begin{figure}[H]
\includegraphics[width=1.0\linewidth,clip]{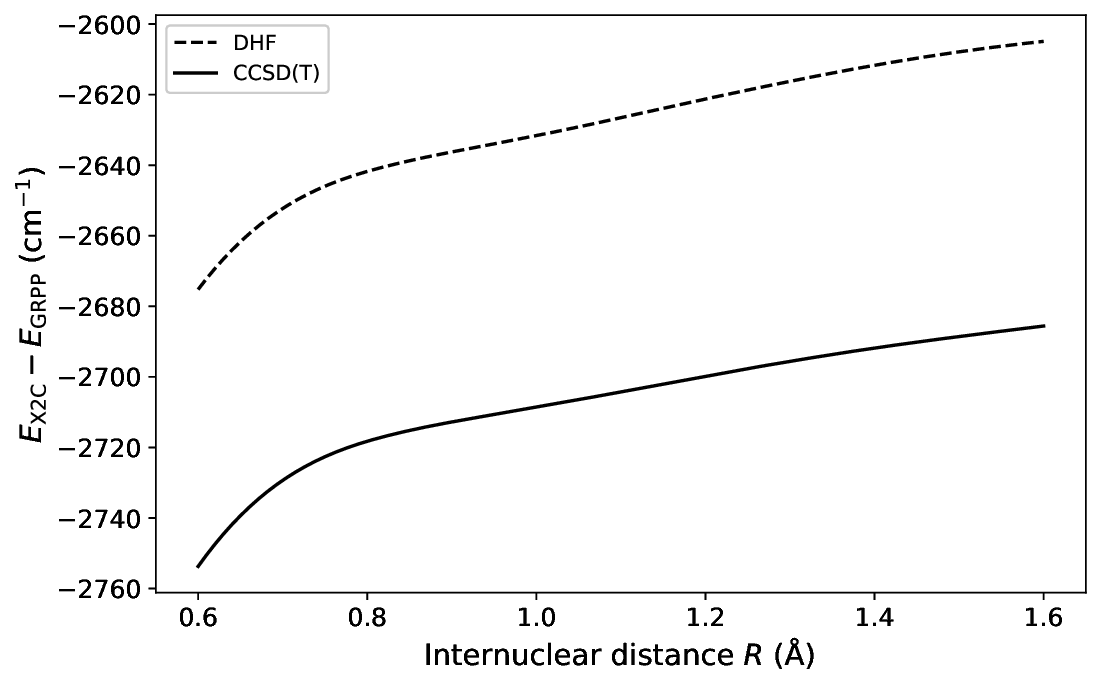}
\caption{The disparities in the PECs computed for the X2C Hamiltonian (including the Gaunt contribution) and the GRPP Hamiltonian. These calculations have been systematically conducted within both the DHF and CCSD(T) approximations.
}\label{Fig:X2C_GRPP}
\end{figure}

The PECs difference obtained for the X2C (with the Gaunt contribution) and GRPP is presented on Fig.~\ref{Fig:X2C_GRPP}. The calculations are performed within the DHF and CCSD(T) approximations. The difference is about $2720$~cm$^{-1}$, but rather stable, within $50$~cm$^{-1}$ for the internulcear distance range. We employ the GRPP approximation to investigate subtle effects such as higher-order correlations or QED corrections, where extremely high accuracy is not required.

As mentioned earlier, the investigation makes use of the standard basis sets from the cc-pV$N$Z family ($N = 4, 5, 6$)~\cite{Dunning:JCP:1989,Kendall:JCP:1992} on both atomic nuclei. Additionally, we employ Dyall's basis sets~\cite{DYALL}, which, after undergoing the extrapolation procedure, yield analogous outcomes.
If not stated otherwise, all the $14$ electrons are considered as the correlated ones, while the virtual states with the energies larger than  $300$~a.u. (which is more than enough) are excluded. 
The exploration of supplementary diffuse functions, which holds particular significance, was also meticulously undertaken, primarily due to their pronounced importance in the accurate evaluation of the dipole moment. In order to gauge the impact, we introduced an augmentation to the existing basis sets by inclusively appending additional diffuse (low exponent) basis functions in a balanced manner.
Basis sets of this nature are designated by the prefix $n$-aug, signifying the inclusion of $n$ supplementary diffuse basis functions within each angular symmetry block. This augmentation strategy serves to systematically assess the implications of such additional functions on the computed outcomes.

\begin{figure}
\includegraphics[width=1.0\linewidth,clip]
{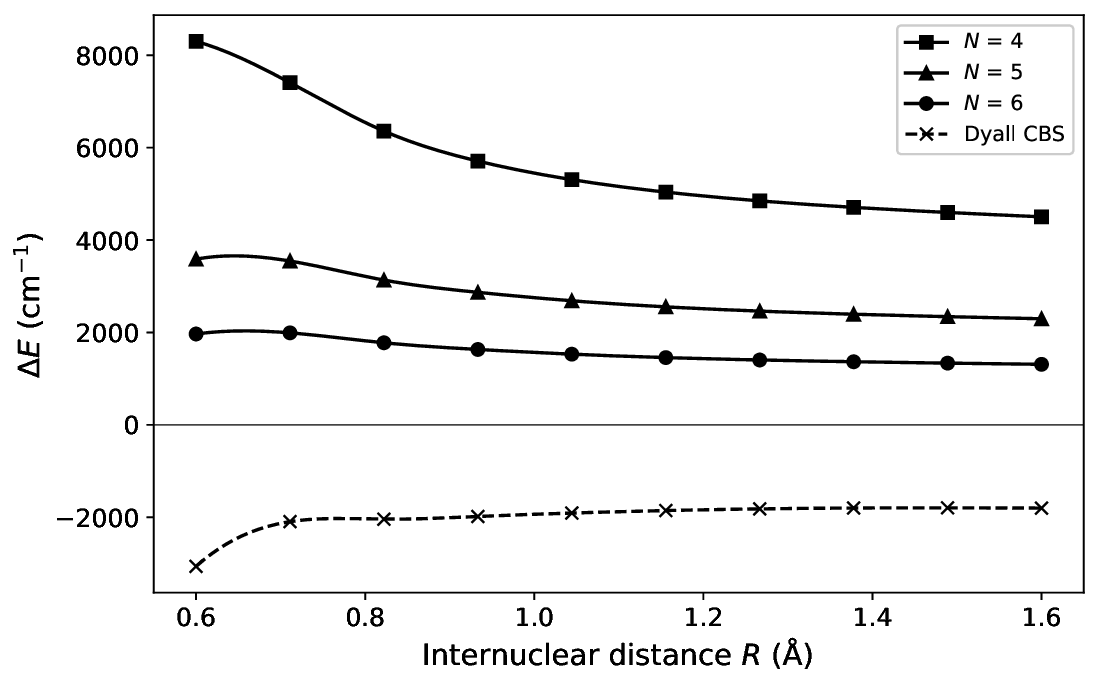}
\caption{Basis set convergence for the ground state PEC. The results are presented concerning the aug-cc-pV$N$Z basis set family relative to the CBS values within the CCSD(T) approximation: $\Delta E = E(\text{aug-cc-pV}N\text{Z}) - E(\text{CBS})$. Additionally, for comparative purposes, the CBS values obtained for the Dyall basis sets are also displayed.}\label{Fig:convergence}
\end{figure}

A demonstrative illustration of the basis convergence analysis conducted under the CCSD(T) approximation is depicted in Figure~\ref{Fig:convergence}. The data set is procured from the aug-cc-pV$N$Z basis set series, wherein the values are rendered relative to the CBS outcome. The measure of convergence error bars, amounting to approximately $2000$~cm$^{-1}$, is derived from the disparity between the outcomes of the most extensive basis set and the CBS values, specifically for the ground state energy.
Given that the focus pertains to the behavior of the PEC under scrutiny, while the absolute magnitudes remain non-crucial, the uncertainty associated with the PEC is gauged by the divergence in behavior between the PECs corresponding to the most expansive basis set and the CBS values, approximating to around $300$~cm$^{-1}$.
As anticipated, the uncertainty pertinent to the relative PEC manifests an improvement by nearly an order of magnitude in comparison to that associated with the absolute values. It's worth noting that convergence is comparatively less favorable for smaller $R$, possibly due to the inherent unoptimization of standard basis sets for cases involving closely spaced atomic nuclei.
For purposes of comparison, the figure includes the CBS value derived from the Dyall basis set family, which exhibits reasonable agreement with the aug-cc-pV$N$Z CBS result.

\par
\begin{figure}
\includegraphics[width=1.0\linewidth,clip]
{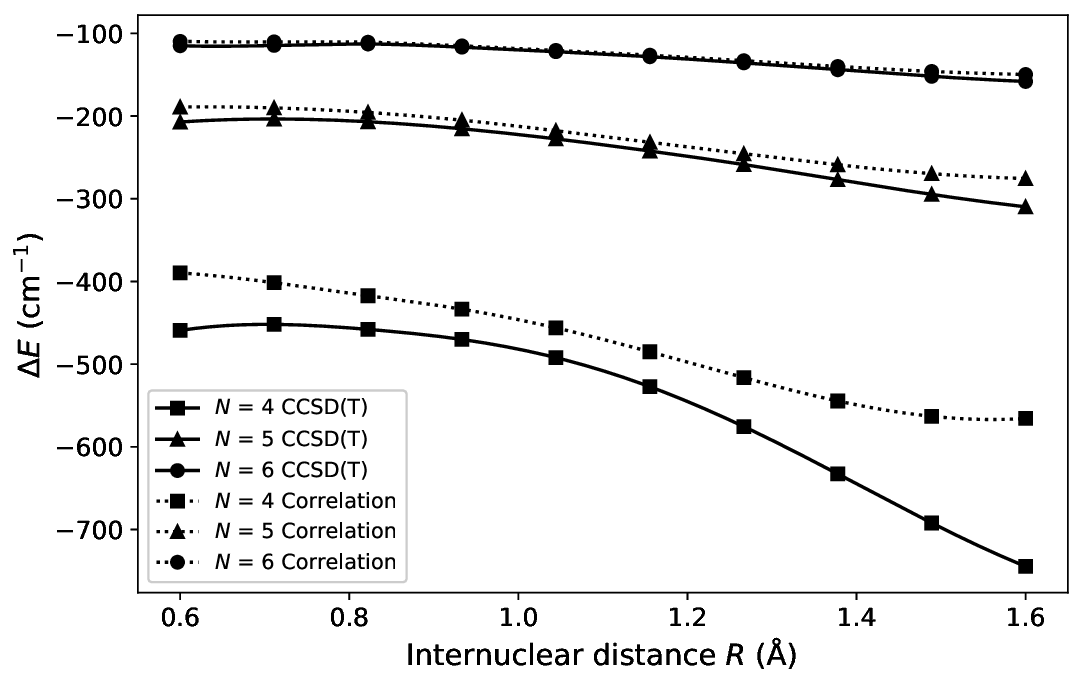}
\caption{Contribution of diffuse orbital basis functions to the PEC of the ground state. The results are presented within the cc-pV$N$Z basis set family, considering both the DHF and CCSD(T) approximations. The "Correlation" notation denotes the $(E_\text{CCSD(T)} - E_\text{DHF})$ contribution.}\label{Fig:diffuse}
\end{figure}

Sensitivity analysis of the ground state PEC concerning the inclusion of diffuse basis functions is exemplified in Figure~\ref{Fig:diffuse}. These investigations encompass calculations conducted within both the DHF and CCSD(T) approximations. The incremental addition of multiple augmentation functions yields results that exhibit minimal deviation. It is worth highlighting that the cc-pV$6$Z basis set, which represents the most extensive among those considered, demonstrates a commendable capability in characterizing correlation effects on its own merit. Nonetheless, it is crucial to underscore the continued relevance of diffuse functions, particularly in enhancing the accuracy of the DHF contribution.

\begin{figure}
\centering
\includegraphics[width=1.0\linewidth,clip]
{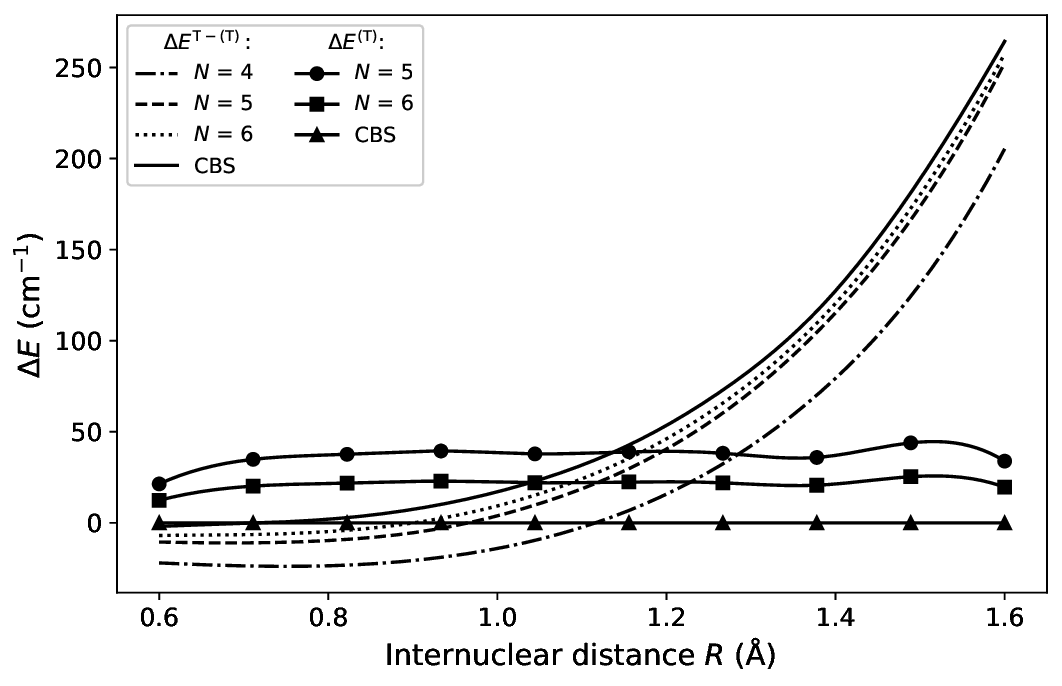}
\caption{\label{Fig:triple}
Basis set convergence of the perturbative triple $\Delta E^\text{(T)}=E_\text{CCSD(T)}-E_\text{CCSD}$ and residual triple $\Delta E^\text{T-(T)}=E_\text{CCSDT}-E_\text{CCSD(T)}$ cluster amplitudes contributions to the PEC of the ground state. The $\Delta E^\text{(T)}$ results are presented relative to the CBS values. These data are obtained under the GRPP approximation and consider explicit correlation of $10$ electrons.}
\end{figure}

In the context of this investigation, we have undertaken the first analysis of the nonperturbative contributions stemming from the triple cluster amplitudes in the case of carbon monoxide, to the best of our knowledge. The outcomes of these calculations, relative to those within the framework of CCSD(T), are shown in Figure~\ref{Fig:triple}. These data are obtained within the GRPP approximation, considering explicit correlation effects among $10$ electrons and excluding the virtual states with the energies larger than $30$~a.u.
Our findings reveal that the residual T cluster amplitude contribution $\Delta E^\text{T-(T)}=E_\text{CCSDT}-E_\text{CCSD(T)}$ is nearly negligible at small $R$ but steadily escalates, ultimately reaching a magnitude of approximately $250$ cm$^{-1}$ at $R$=1.6~\AA. 
\par
\begin{figure}
\centering
\includegraphics[width=1.0\linewidth,clip]
{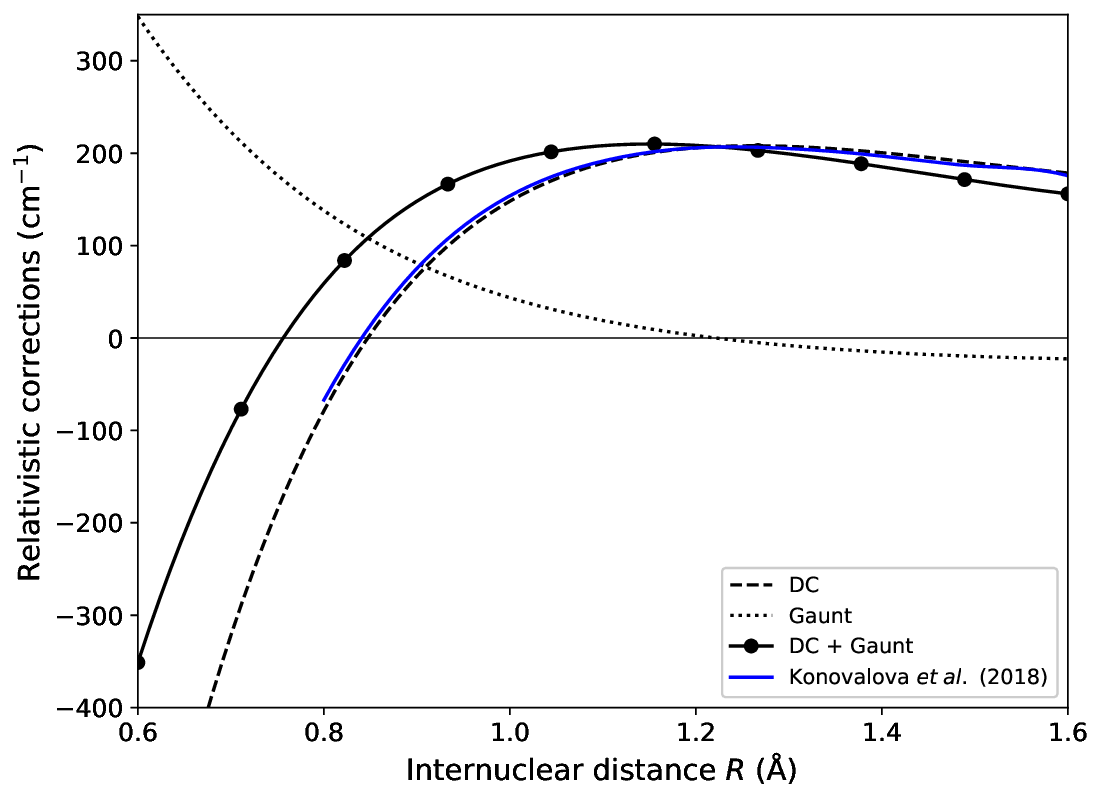}
\caption{\label{Fig:relativistic}
Relativistic corrections to the PEC of the ground state.
The dashed and dotted lines delineate the contributions arising from the DC Hamiltonian and Gaunt corrections, respectively. The solid line represents the cumulative effect. These data are acquired using the CCSD(T) approximation.
The results obtained in Ref.~\cite{Konovalova:OS:2018} are presented with a shift to align them with the DC curve for the purpose of a clear comparison.}
\end{figure}

To gain a more comprehensive perspective on the relativistic corrections, we have divided them into two distinct components: the contributions arising from the DC Hamiltonian and the Gaunt inter-electronic term. The results of these calculations, conducted within the framework of the CCSD(T) approximation, are thoughtfully illustrated in Figure~\ref{Fig:relativistic}. It's essential to note that these data have been shifted to a region of zero reference point at the dissociation limit, facilitating a clearer presentation of their relative behaviors and trends.
As anticipated, both contributions exhibit an increasing in magnitude at smaller internuclear distances, 
and partly compensate each other. An intriguing observation is the presence of extremal points within each of these contributions. Specifically, the contribution from the DC Hamiltonian attains its extreme value near the equilibrium internuclear distance, whereas the Gaunt term reaches its extremum at more substantial internuclear separations. The results obtained in Ref.~\cite{Konovalova:OS:2018} are also presented here. It is worth noting a good agreement with our relativistic DC data, which does not include the Gaunt term. To facilitate a clear comparison, the curve has been shifted accordingly.
\par
\begin{figure}
\centering
\includegraphics[width=1.0\linewidth,clip]
{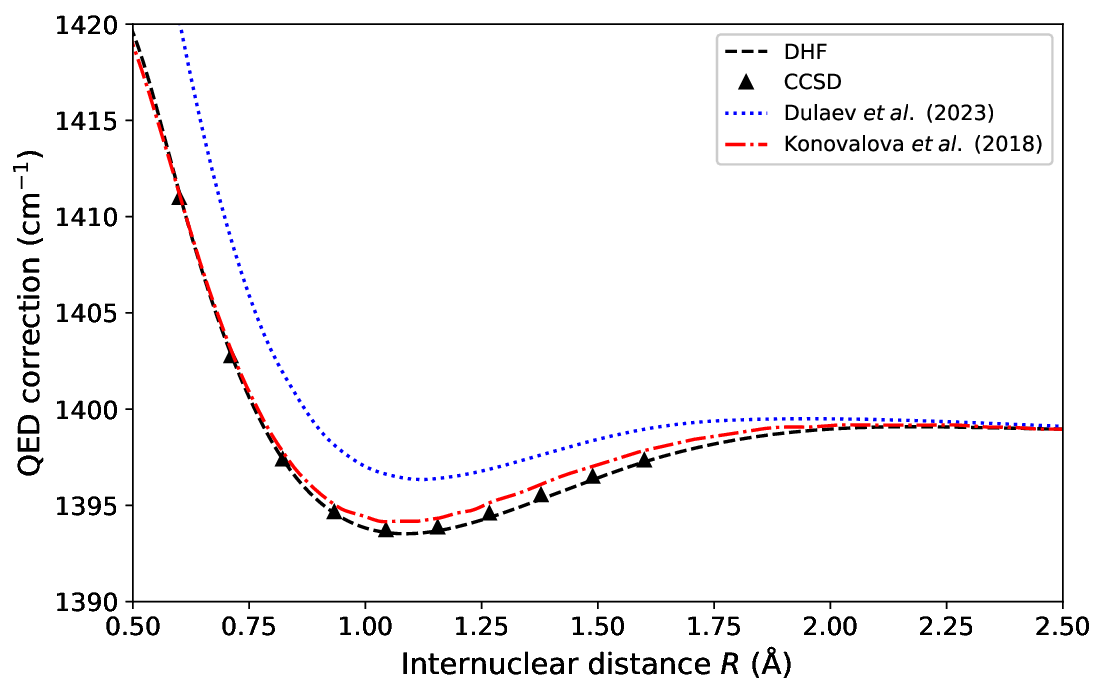}
\caption{\label{Fig:QED}
QED corrections to the PEC of the ground state. The calculations are performed in the DHF and  CCSD approximations. The data obtained by Konovalova \textit{et al.}~\cite{Konovalova:OS:2018} and Dulaev \textit{et al.}~\cite{Dulaev:OS:2023} are also presented.}
\end{figure}

The dependency of the QED correction on the ground state of carbon monoxide concerning internuclear distance is visually depicted in Figure~\ref{Fig:QED}. These calculations have been meticulously conducted within both the DHF and CCSD approximations. A discernible feature is the presence of a minimum within the QED correction, positioned in the vicinity of the equilibrium internuclear distance. This minimum manifests itself at approximately $5$~cm$^{-1}$ concerning its relative position with respect to the dissociation limit. The incorporation of correlation effects engenders a slight (almost negligible) shift of this minimum towards smaller values of $R$. 
An approximate Lamb shift function, semi-empirically estimated~\cite{Konovalova:OS:2018} by scaling the one-electron Darwin correction~\cite{Pyykko:PRA:2001}, is included for comparison. Additionally, the QED correction, recently obtained using the model-QED-operator approach, and configuration interaction method based on Dirac-Sturm orbitals~\cite{Dulaev:OS:2023}, is presented. While the latter approach is more closely related to our method, it's worth noting that our results show a significantly better agreement with the findings from Ref.~\cite{Konovalova:OS:2018}.
\par
\begin{figure}
\centering
\includegraphics[width=1.0\linewidth,clip]
{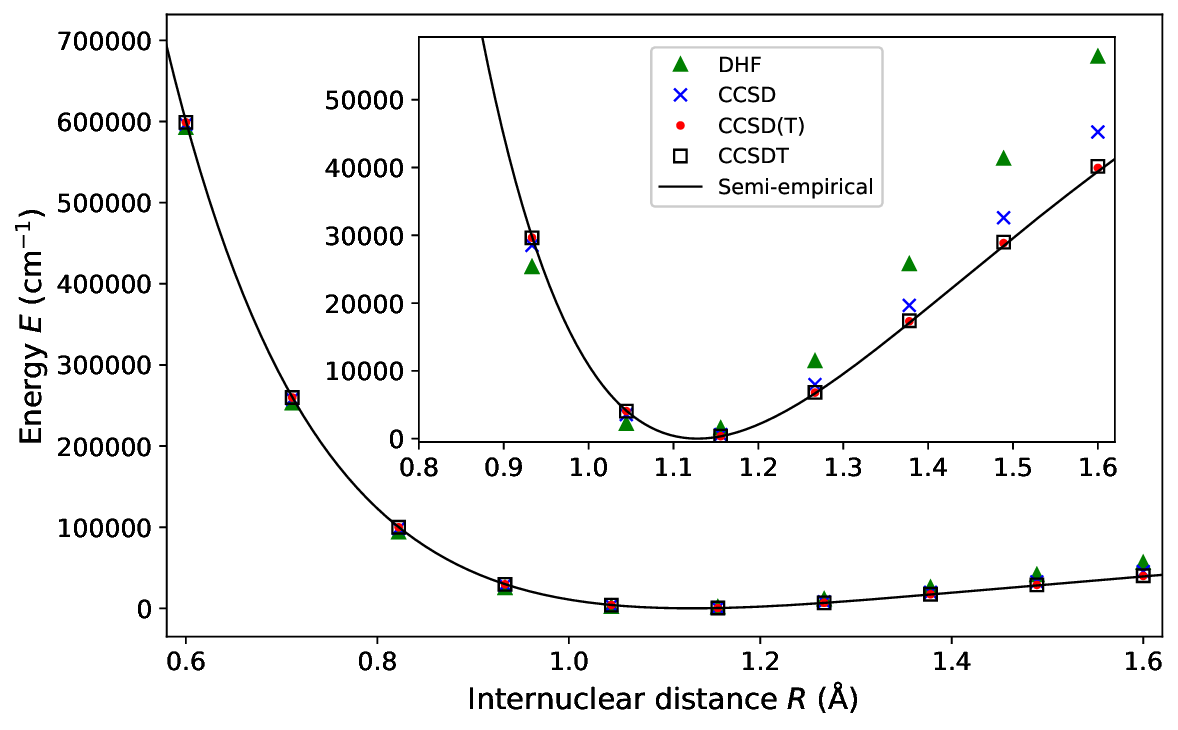}
\caption{\label{Fig:emperical}
The PEC of the ground state obtained at various levels of accounting for correlation effects: DHF, CCSD, CCSD(T), CCSDT, while including all other relevant corrections. The semi-empirical potential curve is extracted from Ref.~\cite{PEF_EMPIRICAL}. The minima of the potential curves are normalized to zero.}
\end{figure}
\begin{figure}
\centering
\includegraphics[width=1.0\linewidth,clip]
{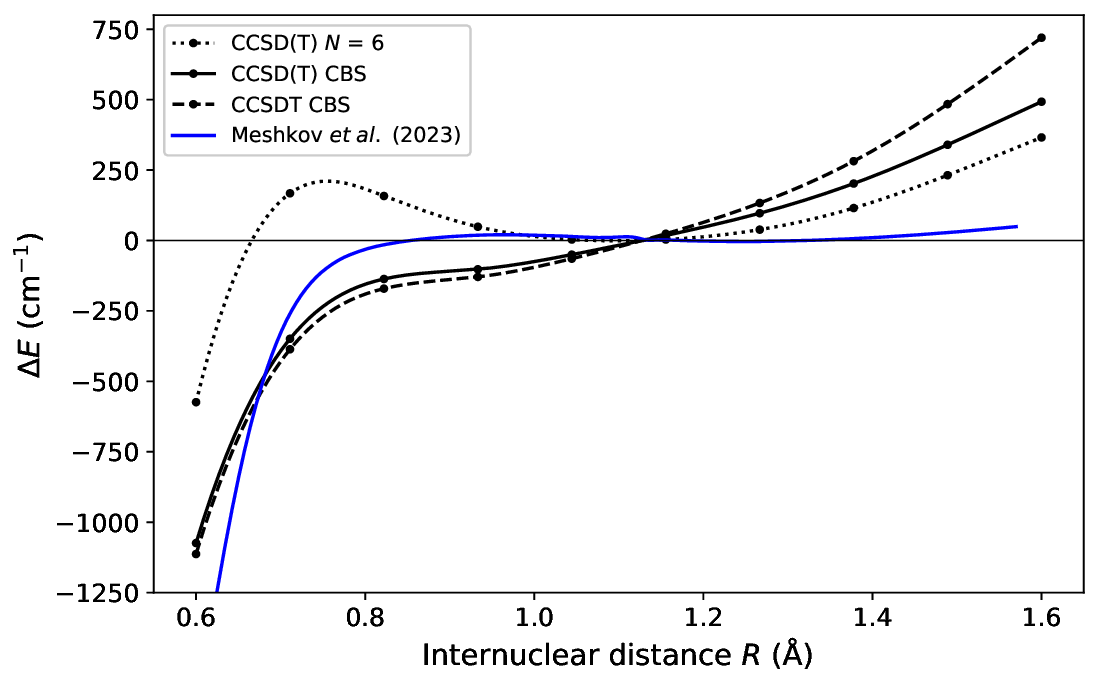}
\caption{\label{Fig:emperical_diff}
Deviation of the CCSD(T) and CCSDT PECs from the semi-empirical potential~\cite{PEF_EMPIRICAL} for the ground state. The results are showcased for calculations conducted using the aug-cc-PV$6$Z basis set and CBS calculations. The relativistic and QED corrections are taken into account. Additionally, for comparative purposes, the data obtained by Mehskov \textit{et al.} Ref.~\cite{Meshkov:RJPC:2023} obtained within the MR-ACPF approach are also displayed. 
}
\end{figure}

Finally, in Figure~\ref{Fig:emperical}, we present a comparative analysis of our computed results with a semi-empirical potential~\cite{PEF_EMPIRICAL}. The PECs correspond to various levels of considering correlation effects, including DHF, CCSD, CCSD(T), CCSDT, while accounting for all other relevant corrections. To facilitate a meaningful comparison, these PECs and the semi-empirical potential are aligned to zero at their respective minimum points. Within the vicinity of the minimum, a notable agreement between the approaches is observed. However, as the internuclear distance increases, distinctions from the CCSDT results become increasingly pronounced, reaching approximately $17000$ and $6000$~cm$^{-1}$ concerning deviations from the DHF and CCSD outcomes, respectively.
Further insight into the deviations of the CCSD(T) and CCSDT curves from the semi-empirical potential for the ground state is presented in Figure~\ref{Fig:emperical_diff}. To ensure a meaningful comparison, these curves are synchronized at the equilibrium points $R_{\rm e}\approx 1.128$~\AA, with the inclusion of relativistic and QED corrections. It is important to note that the residual triple amplitude contributions primarily come into play at larger $R$. The uncertainty associated with the PEC is predominantly dictated by the basis set convergence in the CCSD(T) calculations and is estimated to be approximately $300$ cm$^{-1}$.
For purposes of comparison, the figure includes the value recently derived by Meshkov \textit{et al.} within the averaged coupled pair functional (ACPF) approach~\cite{Meshkov:RJPC:2023}.

Furthermore, our investigation was extended to an examination of the bond distance's dependence on various facets of the calculations and the associated approximations. In pursuit of precision, additional evaluations in proximity to the bond distance were conducted. The outcome of this scrutiny regarding basis set convergence within the cc-pV$N$Z basis set series is presented in Table~\ref{Table:Bond_distance}. Notably, the accuracy of the CBS value is estimated to be within the order of $1$~m\AA. Furthermore, it is observed that the influence of diffuse functions on the optimal geometry is minimal. Interestingly, the residual triple and Gaunt contributions exhibit a compensatory effect, collectively effecting a correction of less than $0.1$~m\AA ~to the final value.
\onecolumngrid\
\begin{table}[H]
\centering
\caption{Basis set convergence for the equilibrium bond distance $R_{\rm e}$ (in \AA) of the ground CO state. The results are presented for the cc-pV$N$Z basis set family within the CCSD, CCSD(T) and CCSDT approximations. Additionally, the impact of the Gaunt term correction noted as "+G" is also demonstrated. The semi-empirical $R_{\rm e}$=1.128217 (\AA) value is adopted from Ref.~\cite{R_emp}.}
\begin{tabular}{lllll}
\hline\hline
$N$    & CCSD & CCSD(T) & CCSD(T)+G & CCSDT+G    \\ 
\hline
4  & 1.1225 & 1.1297      & 1.1299      & 1.1297 \\
5  & 1.1213 & 1.1285      & 1.1287      & 1.1285 \\
6  & 1.1208 & 1.1280      & 1.1282      & 1.1280 \\
\hline
CBS & 1.1202 & 1.1274      & 1.1276      & 1.1274 \\
\hline
\end{tabular}
\label{Table:Bond_distance}
\end{table}
\twocolumngrid\

\begin{figure}
\centering
\includegraphics[width=1.0\linewidth,clip]{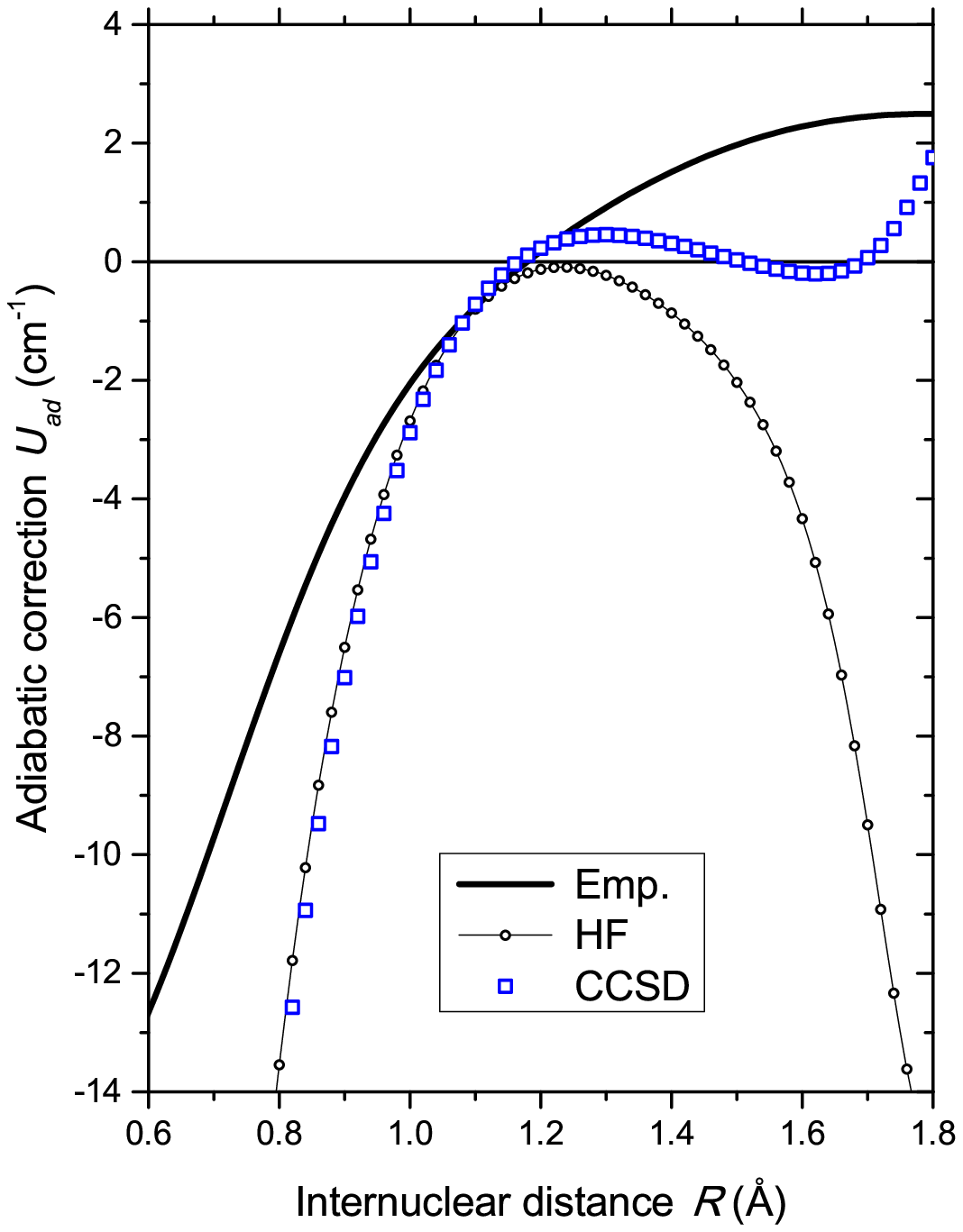}
\caption{\label{Fig:DBOC} Comparison of the mass-dependant DBOC functions for the ground-state of $^{12}$C$^{16}$O molecule: the HF and CCSD methods versus empirical data from Ref.~\cite{PEF_EMPIRICAL}.}
\end{figure}

The DBOC calculated for the ground state of the most abundant isotopologue $^{12}$C$^{16}$O under the HF and CCSD approximations are compared on Figure~\ref{Fig:DBOC} with the relevant adiabatic correction derived empirically in Ref.~\cite{PEF_EMPIRICAL}. The original {\it ab initio} functions were vertically shifted to match their pure empirical counterpart at the point of equilibrium distance $R_{\rm e}$. The difference of the HF and CCSD curves clearly observed at $R>1.2$ (\AA) highlights the rapid increase of the electronic correlation effect as $R$ increases.  

\begin{figure}
\centering
\includegraphics[width=1.0\linewidth,clip]{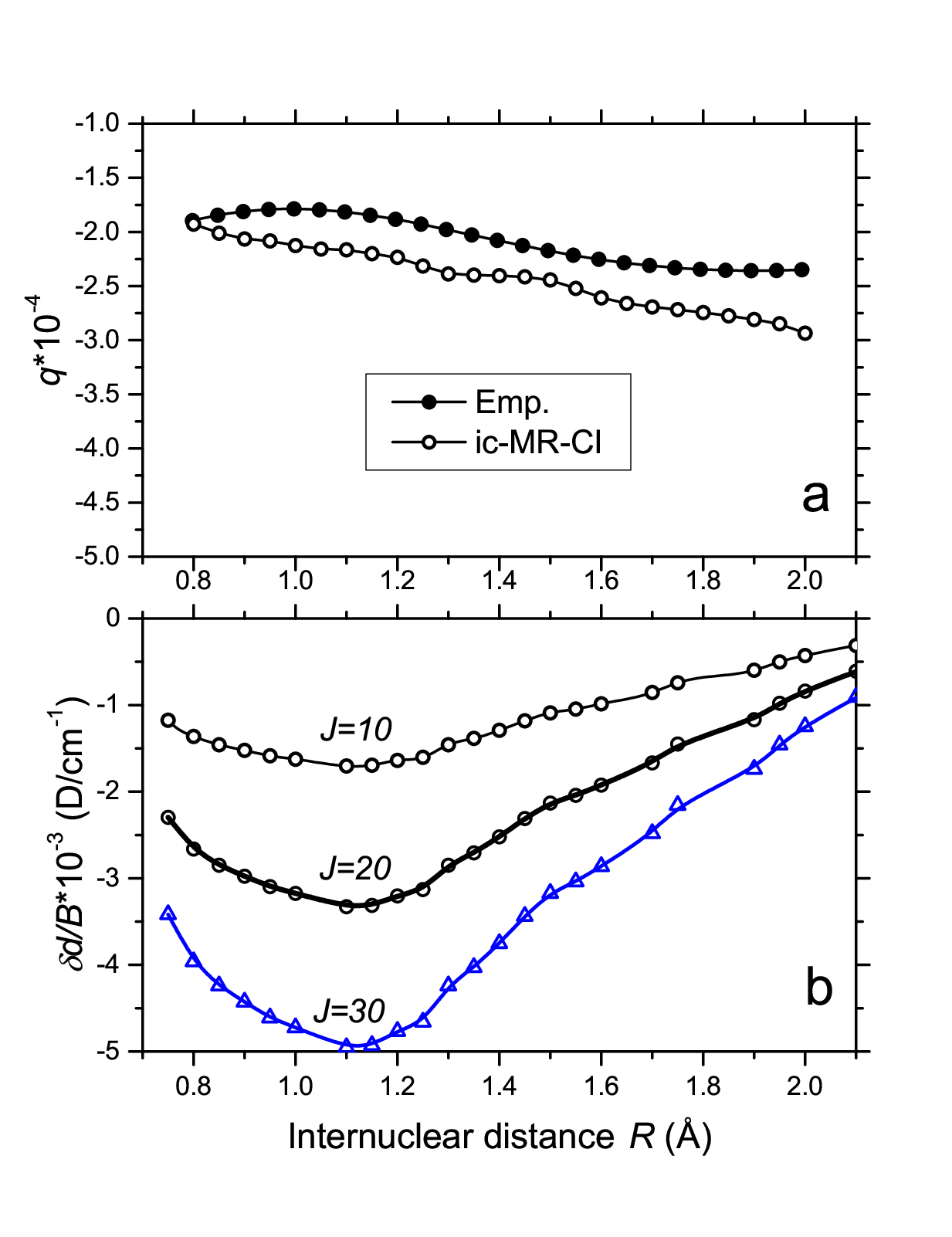}
\caption{\label{Fig:q}
({\bf a}) Comparison of the dimensionless mass-dependent $q$-correction, evaluated according to Eq.~(\ref{hetU}), to rotational energy of the $^{12}$C$^{16}$O ground-state with its empirical counterpart~\cite{PEF_EMPIRICAL}. 
({\bf b}) The $J$-dependent contribution of the non-adiabatic effect into the permanent dipole function for the CO ground-state.}
\end{figure}

Moreover, as can be seen on Figure~\ref{Fig:q}a the 2-nd order PT correction to a rotational part of the effective mass-dependant potential for the ground-state of the $^{12}$C$^{16}$O isotopologue, which was estimated using the ic-MR-CI calculations, agrees very well with its empirical counterpart from Ref.~\cite{PEF_EMPIRICAL}.

\begin{figure}
\centering
\includegraphics[width=1.0\linewidth,clip]
{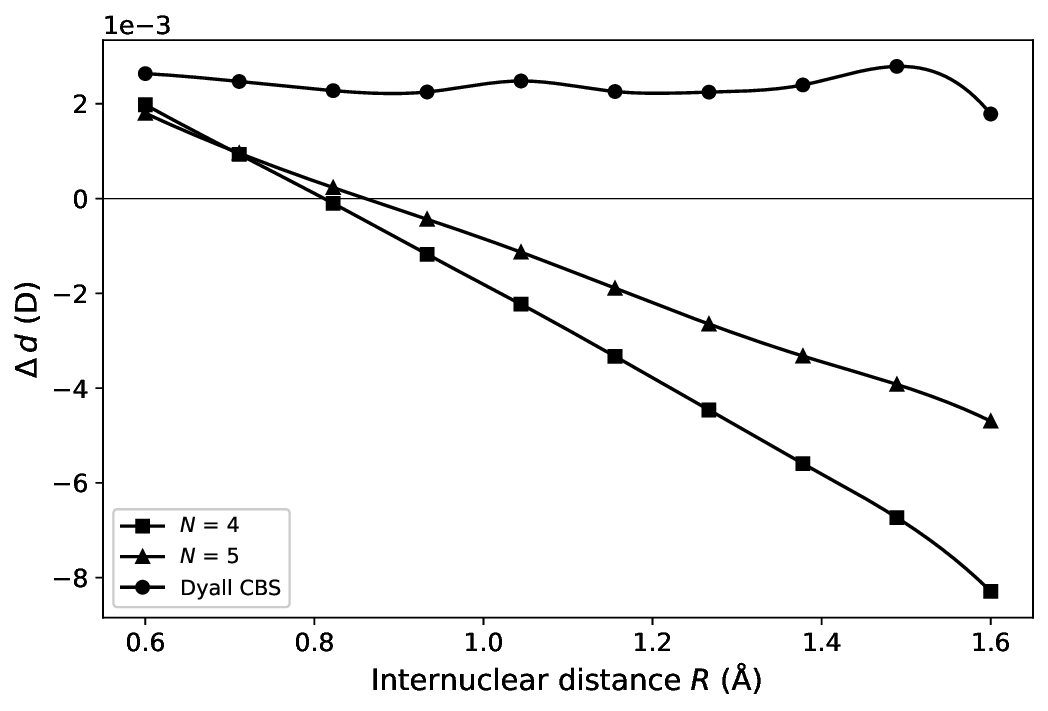}
\caption{\label{Fig:DM_convergence}
Basis set convergence for the DMC of the
ground state. The results are showcased concerning the
aug-cc-pV$N$Z basis set family relative to the CBS
values within the CCSD(T) approximation: $\Delta \mu = \mu(\text{aug-cc-pV}N\text{Z}) - \mu(\text{CBS})$. For comparative purposes, the CBS values obtained for the Dyall basis sets are also presented.
}
\end{figure}

\begin{figure}
\centering
\includegraphics[width=1.0\linewidth,clip]
{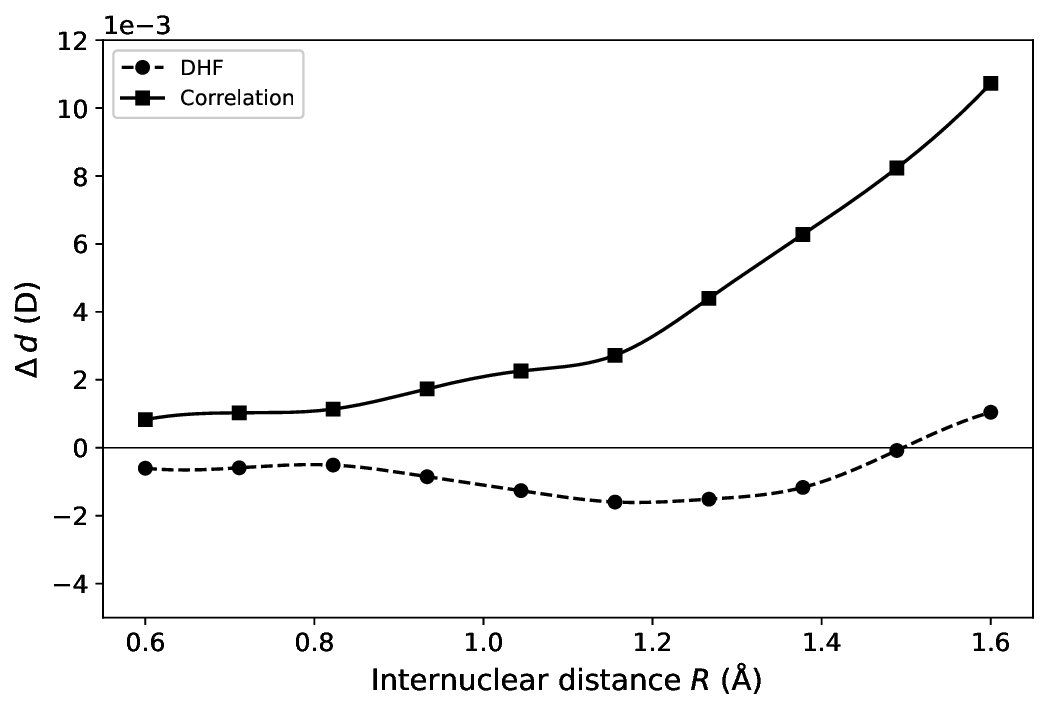}
\caption{\label{Fig:DM_diffuse}
Contribution of diffuse orbital basis functions to the ground-state DMC. The results are illustrated for the cc-pV$5$Z basis set family within both DHF and CCSD(T) approximations.
}
\end{figure}

\begin{figure}
\centering
\includegraphics[width=1.0\linewidth,clip]
{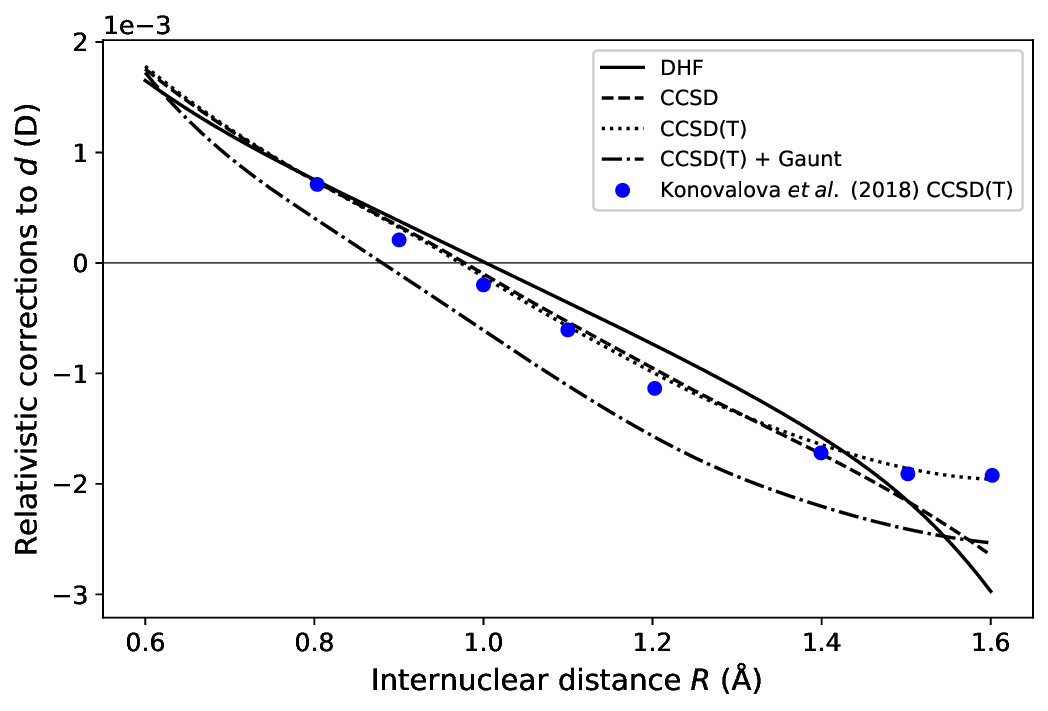}
\caption{\label{Fig:DM_relativistic}
Relativistic correction to the ground-state DMC. 
The calculations are executed within the DHF, CCSD and CCSD(T) approximations. Additionally, for comparative purposes, the data obtained by Konovalova \textit{et al.} Ref.~\cite{Konovalova:OS:2018} are also displayed.
}
\end{figure}

A parallel analysis has been conducted for the DMC of the ground state of carbon monoxide. The investigation centered on basis set convergence, utilizing the aug-cc-pV$N$Z basis set series, is graphically depicted in Figure~\ref{Fig:DM_convergence}. 
Through an overarching assessment and a comparative evaluation with results obtained using the Dyall basis sets, we estimate the associated uncertainty to be within the range of $3$~mD. Similar to the ground energy scenario, it is noteworthy that the inclusion of just one diffuse orbital basis function per symmetry proves to be adequate for attaining DMC accuracy at the level of approximately $1$~mD. The individual contributions pertaining to these aspects are elucidated in Figure~\ref{Fig:DM_diffuse}.
Delving into the relativistic correction to the DMC, as delineated in Figure~\ref{Fig:DM_relativistic}, we observe that this contribution amounts to a few mD, while the Gaunt contribution registers an order of magnitude smaller. These calculations have been conducted within the DHF, CCSD, and CCSD(T) approximations. The CCSD(T) ones without the Gaunt contribution are in a reasonable agreement with the Ref.~\cite{Konovalova:OS:2018}.
The rotational part of non-adiabatic correction to the DMC function, as predicted by Eq.~(\ref{hetd}) for the $^{12}$C$^{16}$O isotopologue, is illustrated in Figure~\ref{Fig:q}b for different $J$-values. The absolute magnitude of this mass-dependent correction is comparable to the relativistic correction to the DMC (see Figure~\ref{Fig:DM_relativistic}). The {\it ab initio} ic-MR-CI calculations revival that the dominant perturbation is caused by the lowest excited $A^1\Pi$ state.

\begin{figure}
\centering
\includegraphics[width=1.0\linewidth,clip]
{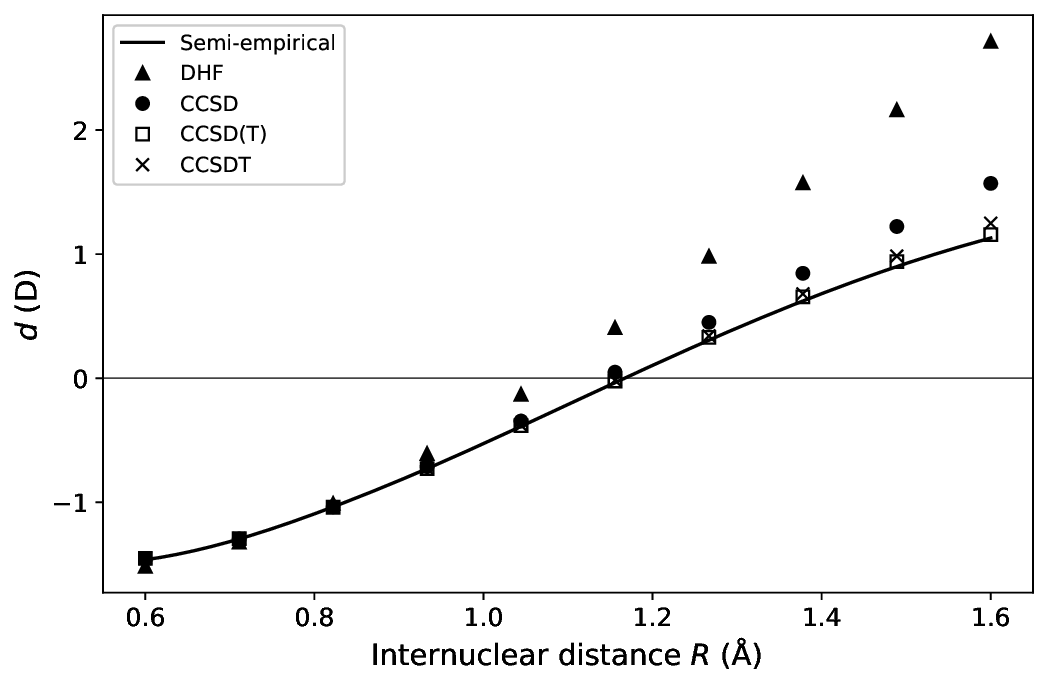}
\caption{\label{Fig:DM_empirical}
The ground-state DMC obtained at various levels of accounting for correlation effects: DHF, CCSD, CCSD(T), CCSDT, while incorporating all other relevant corrections. The semi-empirical curve is extracted from Ref.~\cite{Meshkov:JQSRT:2022}.
}
\end{figure}

\begin{figure}
\centering
\includegraphics[width=1.0\linewidth,clip]
{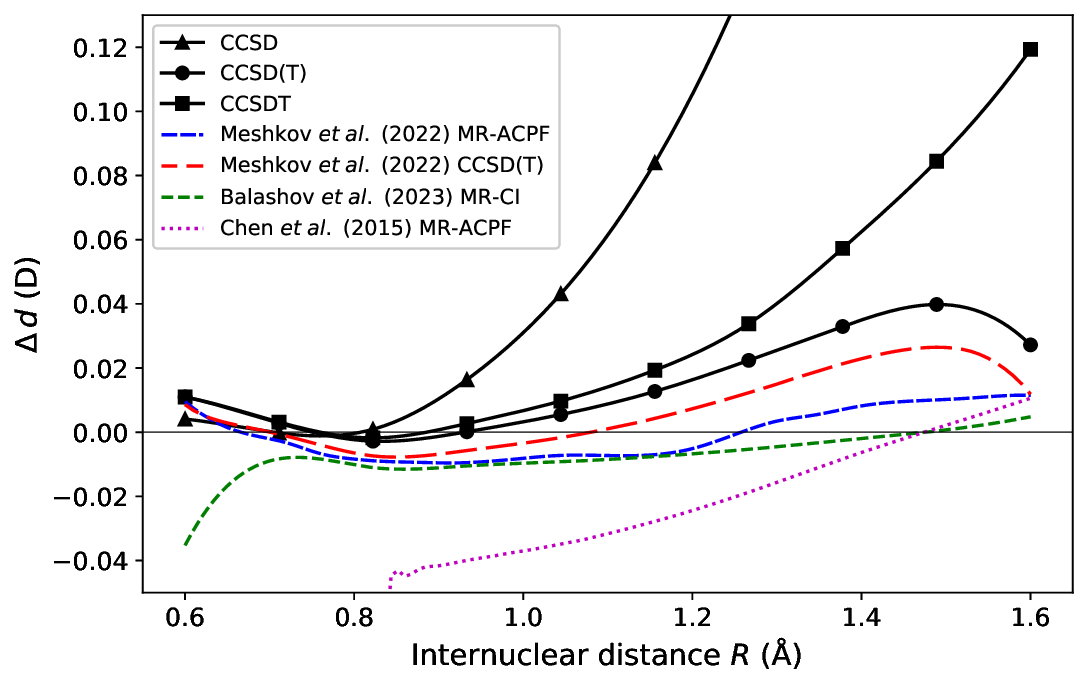}
\caption{\label{Fig:DMF_empirical_diff}
Deviation of the DMC obtained using the CCSD, CCSD(T), and CCSDT methods from the semi-empirical curve~\cite{Meshkov:JQSRT:2022} for the ground state. The results of 
MR-ACPF and CCSD(T) calculations by Meshkov \textit{et al.}~\cite{Meshkov:JQSRT:2022},
MR-CISD(+Q)-FF by Balashov \textit{et al.}~\cite{Balashov:JCP:2023} and
MR-ACPF by Chen \textit{et al.}~\cite{Chen:CPB:2015} 
are also presented. 
}
\end{figure}

Finally, in Figure~\ref{Fig:DM_empirical}, we present a comparative analysis of our computed DMCs with a semi-empirical curve~\cite{Meshkov:JQSRT:2022}.
Within the realm of smaller internuclear distances, rather good agreement among the different approaches is observed. However, as internuclear distances increase, the influence of correlation effects becomes notably substantial. To provide a detailed perspective, we have illustrated the deviations of the CCSD, CCSD(T), and CCSDT DMCs from the semi-empirical curve in Figure~\ref{Fig:DMF_empirical_diff}.
The accuracy of the evaluated DMCs falls within the range of $5$~mD and is primarily determined by basis set convergence. Nevertheless, it is pertinent to note that the dipole moment contributions stemming from the triple cluster amplitudes, both perturbative and residual, exhibit noteworthy magnitudes of up to $0.5$ and $0.1$ debye, respectively, particularly at larger internuclear distances. It is plausible that the deviation of CCSDT results from the semi-empirical values is attributed to the contributions from higher-order excitation. The results of 
MR-ACPF calculations by Chen \textit{et al.}~\cite{Chen:CPB:2015}, 
MR-ACPF and CCSD(T) by Meshkov \textit{et al.}~\cite{Meshkov:JQSRT:2022} and
multi-reference CI (MR-CI(+Q)-FF) by Balashov \textit{et al.}~\cite{Balashov:JCP:2023}, 
are also presented for comparison.

\section{Summary and Conclusions}\label{sec:concl}
%
The ground-state potential energy and dipole moment profiles of carbon monoxide have been rigorously examined utilizing \textit{ab initio} methods within the framework of the relativistic coupled-cluster approach. This study incorporates non-perturbative single, double, and triple cluster amplitudes, complemented by a finite-field methodology. The nonperturbative triple-amplitude corrections were evaluated for the first time. The calculations were conducted using the implementation described in the references \cite{Oleynichenko:Sym:2020,Oleynichenko:2020,Oleynichenko:website}.
The generalized relativistic pseudo-potential model was employed to effectively introduce relativistic effects into the all-electron correlation treatment and to account for quantum-electrodynamics (QED) corrections through the model-QED-operator approach.

We conducted a detailed investigation into the sensitivity of the results to the parameters of the basis set and approximations employed. Our findings are in satisfactory agreement with the most precise available semi-empirical data.

For further enhancement of calculation accuracy and the potential extension of the range of considered internuclear distances, particularly towards larger distances, it appears necessary to employ more computationally demanding multiconfigurational variants of the coupled-cluster method (and/or configuration interaction). These approaches would adequately account for the significant increase in the multiconfigurational character of the electronic wave function of the CO ground state at intermediate and, especially, large internulcear distances.
\subsection*{\bf Acknowledgments}
We are grateful to L.V.~Skripnikov and A.V.~Oleynichenko for helpful discussions.
The study was supported by the Russian Science Foundation (interdisciplinary grant No. 22-62-00004).
%
%
\newpage
\bibliographystyle{apsrev4-2}
\bibliography{bibliography}


\end{document}